\documentclass[%
reprint,
superscriptaddress,
amsmath,
amssymb,
aps,
pra,
longbibliography
]{revtex4-2}

\usepackage[T1]{fontenc}
\usepackage[utf8]{inputenc}

\usepackage{times}
\usepackage[unicode=true,breaklinks=true,colorlinks=true]{hyperref}
\hypersetup{citecolor=blue,urlcolor=blue}
\usepackage{graphicx}
\usepackage{bm}
\usepackage{bbold}

\usepackage{mathtools}
\usepackage{tensor}

\begin{document}
\title{Precise spectroscopy of high-frequency oscillating fields with a single qubit sensor}
\author{Yaoming Chu}
\author{Pengcheng Yang}
\email{pengchengyang@hust.edu.cn}
\affiliation{School of Physics, Institute for Quantum Science and Engineering, Huazhong University of Science and Technology, Wuhan 430074, China}
\affiliation{International Joint Laboratory on Quantum Sensing and Quantum Metrology, Huazhong University of Science and Technology, Wuhan 430074, China}
\affiliation{Wuhan National Laboratory for Optoelectronics, Huazhong University of Science and Technology, Wuhan 430074, China}
\affiliation{State Key Laboratory of Precision Spectroscopy, East China Normal University, Shanghai, 200062, China}
\author{Musang Gong}
\author{Min Yu}
\author{Baiyi Yu}
\affiliation{School of Physics, Institute for Quantum Science and Engineering, Huazhong University of Science and Technology, Wuhan 430074, China}
\affiliation{International Joint Laboratory on Quantum Sensing and Quantum Metrology, Huazhong University of Science and Technology, Wuhan 430074, China}
\author{Martin B. Plenio}
\affiliation{Institut f\"{u}r Theoretische Physik und IQST, Universit\"{a}t Ulm,
Albert-Einstein-Allee 11, D-89069 Ulm, Germany}
\affiliation{International Joint Laboratory on Quantum Sensing and Quantum Metrology, Huazhong University of Science and Technology, Wuhan 430074, China}
\author{Alex Retzker}
\affiliation{Racah Institute of Physics, The Hebrew University of Jerusalem, Jerusalem 91904, Givat Ram, Israel}
\author{Jianming Cai}
\affiliation{School of Physics, Institute for Quantum Science and Engineering, Huazhong University of Science and Technology, Wuhan 430074, China}
\affiliation{International Joint Laboratory on Quantum Sensing and Quantum Metrology, Huazhong University of Science and Technology, Wuhan 430074, China}
\affiliation{Wuhan National Laboratory for Optoelectronics, Huazhong University of Science and Technology, Wuhan 430074, China}
\affiliation{State Key Laboratory of Precision Spectroscopy, East China Normal University, Shanghai, 200062, China}
\date{\today}

\begin{abstract}
Precise spectroscopy of oscillating fields plays significant roles in many fields. Here, we propose an experimentally feasible scheme to measure the frequency of a fast-oscillating field using a single-qubit sensor. By invoking a stable classical clock, the signal phase correlations between successive measurements enable us to extract the target frequency with extremely high precision. In addition, we integrate dynamical decoupling technique into the framework to suppress the influence of slow environmental noise. Our framework is feasible with a variety of atomic and single solid-state-spin systems within the state-of-the-art experimental capabilities as a versatile tool for quantum spectroscopy.
\end{abstract}
\maketitle

\section{Introduction}

Sensors working in quantum regime with new capabilities, large bandwidths, extremely high spatial and spectral resolutions attract increasingly interest in the field of precision metrology \cite{Ramsey1950,Degen2017,Acin2018,DeMille2017,Budker2007,Taylor2008,Casola2018,Aslam2017,Hatridge2011,Cai2014,Moser2013,Kotler2011,Muller2014,Campbell2017,Mccormick2019,Gefen2018,Liu2019,Rotem2019,Cohen2019}. Frequency spectroscopy with single-qubit probes (e.g., nitrogen-vacancy centers in diamond) as one major branch in this field has achieved a significant breakthrough with the development of the quantum heterodyne (Qdyne) technique \cite{Boss2017,Schmitt2017,Glenn2018}. It enables a quantum sensor for precise spectroscopy to go beyond its coherence time by nonlinearly mixing a target signal field with a stable local oscillator \cite{Schmitt2017}.
In particular, precise frequency determination of fast-oscillating fields can have a wide range of significant applications, e.g. high-resolution microwave field spectrum analyzation \cite{Chipaux2015,Shao2016,Horsley2018}, detection of electron spin motions in solids \cite{Sushkov2014,Kolkowitz2015,Hall2016} and nuclear magnetic resonance in the high magnetic field regime \cite{Casanova2018,Aharon2019}. Utilizing the large energy gap to sense high frequency fields was proposed in \cite{Stark2017,Joas2017}, but in these proposals frequency resolution was still limited by the coherence time of the sensor. Conventional quantum lock-in detection, which addresses this problem allows a probe to cumulatively sense an oscillating signal, usually requires implementing a sequence of periodic spin-flipping $\pi$-pulses, the repetition rate of which should be resonant with the target field \cite{Khodjasteh2005,Uhrig2007,Gordon2008,Biercuk2009,Hall2010,Lange2010,Lange2011,Yang2011,Souza2011,Farfurnik2015}. However, the accessible time duration to implement a sharp $\pi$ pulse in experiments is not infinitesimal but a finite interval e.g. a few tens of nanoseconds in solid-state-spin systems, due to the power limitations of the control fields \cite{Stark2017,Aharon2019} and the deleterious high-power heating effects, e.g. in biological environments \cite{Cao2020}, which makes quantum heterodyne detection of high-frequency oscillating fields a big challenge.

\begin{figure}[t]
  \includegraphics[width=84mm]{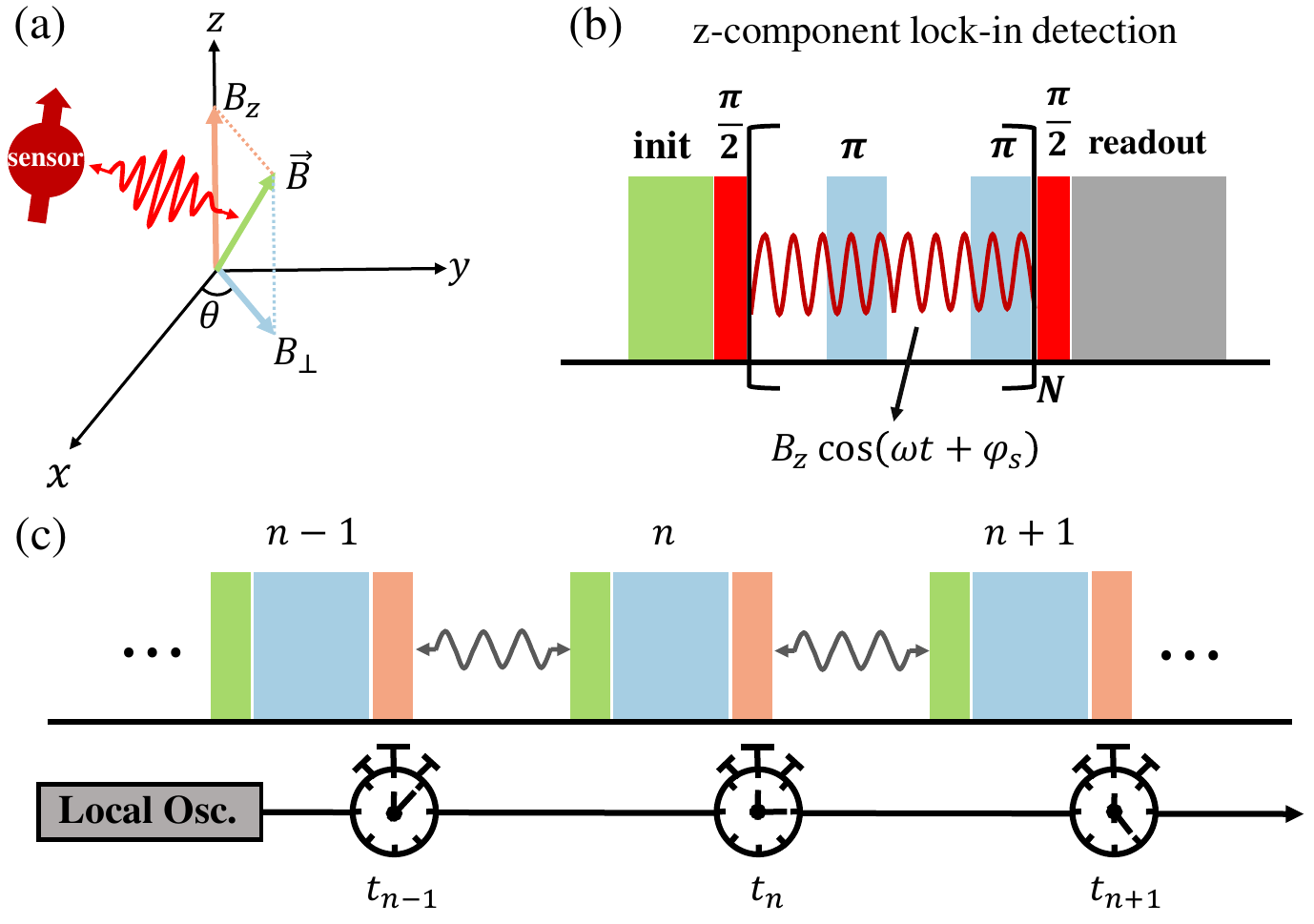}
  \caption{Frequency measurement with a single-qubit probe. (a) A single qubit acting as a probe interacts with an oscillating field. (b) Conventional quantum lock-in detection to measure the longitudinal component of the target signal. (c)  Qdyne measurement chain. Each single measurement consists of initialization (green rectangles) and readout (orange rectangles) of the sensor, interaction with the target field (blue rectangles) and an additional delay time (gray wave lines with arrows in both ends) accompanied by a stable local oscillator for precise time recording.} \label{fig:model}
\end{figure}

In this work, we address this challenge and propose an experimentally feasible scheme for measuring the frequency of a fast-oscillating field using a single qubit sensor based on quantum heterodyne detection. In each single measurement, we effectively obtain a much slower radio-frequency (RF) signal field in a rotating frame, which is then spectrally separated from its noise environment and measured using a dynamical decoupling sequence. Extremely sharp $\pi$-pulses to be synchronized with the fast-oscillating field are thus not required and the sensor's working region can be extended to the high-frequency range under ambient conditions. We further extract information on the target frequency by harnessing phase correlations between successive single measurements based on the Qdyne technique. The proposal is feasible with a solid-state-spin system formed by negatively charged nitrogen-vacancy (NV) center in diamond as well as other atomic qubit systems, and thus serves as a versatile tool for high-resolution quantum spectroscopy.

\section{Effective RF signal in the transverse plane}

We consider a generic two-level system acting as a quantum sensor described by the Hamiltonian $H_p=(\omega_0/2)\sigma_z$, with the Pauli operator $\sigma_z=|0\rangle\langle0|-|1\rangle\langle1|$. The sensor is exposed to a fast-oscillating field, see Fig.\ref{fig:model} (a), with the interaction Hamiltonian assumed as $H_s=\vec{b}\cdot\vec{\sigma}\cos(\omega t+\varphi_s)$ and $\vec{\sigma}=(\sigma_x,\sigma_y,\sigma_z)$, where $\omega$ is the target frequency to be estimated and $\vec{b}=(b_x,b_y,b_z)$ is a vector of signal coupling strength. Conventional quantum lock-in detection usually measures the longitudinal field component of the target field, i.e. $z$-component, which is parallel to the quantization axis of the probe \cite{Ramsey1950,Kotler2011, Schmitt2017}, which however becomes invalid in the case of a fast-oscillating field. It can be seen from Fig.\ref{fig:model} (b) that experimentally accessible finite-width $\pi$ pulses fail to flip the probe spin fast enough to be resonant with the target field. As a result, actions on the probe carrying information about the target frequency can not coherently adds up.

In order to solve this issue the bare energy gap is used to set the probed frequency via the $x$-$y$ plane\cite{Chipaux2015}. As the bare states suffer from short coherence time dynamical decoupling was proved to be efficient in this regime\cite{Stark2017,Joas2017}.
In this work we combine these methods with Qdyne with the aim to increase the resolution at high frequency. We measure transverse field components in the $x$-$y$ plane by tuning the energy splitting $\omega_0$ of the sensor close to the target frequency $\omega$. We remark that such a theoretical setting is feasible with a variety of experimental platforms, e.g. the ground-state energy levels of the NV center electron spin which are split by a few gigahertz due to zero-field splitting and can be adjusted by an external static magnetic field \cite{Taylor2008}. Thus, we tune the difference $\Delta=\omega-\omega_0$ to be on the order of megahertz and make it satisfy $k_s=(b_x^2+b_y^2)^{1/2}\ll |\Delta|\ll \omega+\omega_0$, which results in an effective RF signal in the interaction picture with respect to $H_p=(\omega_0/2)\sigma_z$ as follows (Appendix \ref{appendix-a})
\begin{equation}
  \label{eq:H_RF}
  H_I \approx \frac{k_s}{2}\left[\cos(\Delta t+\varphi)\sigma_x+\sin(\Delta t+\varphi)\sigma_y\right],
\end{equation}
wherein $\varphi=\varphi_s+\theta$, $\tan(\theta)=b_y/b_x$. In order to extract phase information on $\varphi$ from this RF signal and suppress slow environmental noise, a dynamical decoupling sequence nearly resonant with $\Delta$ can be exploited. We stress that the influence of the oscillating field that is applied throughout the entire measurement is negligible in the stage of sensor initialization and readout under the condition $|\Delta|\gg k_s$.

As an illustrative example, we apply a train of  Carr-Purcell-Meiboom-Gill (CPMG) sequences \cite{Carr1954,Meiboom1958,Gullion1990}, i.e. $(\tau$-$\pi_x$-$\tau$-$\pi_x)^{N_s}$, which periodically rotates the sensor around $\hat{x}$ direction. Consequently, we are left in the toggling frame with an effective Hamiltonian
\begin{equation}
  \label{eq:H_CPMG}
  \mathcal{H}(\varphi)\approx \frac{k_s}{\pi}\cos(\delta t+\varphi)\sigma_y,
\end{equation}
where $\delta=\Delta-\pi/\tau$ denotes a small detuning due to lack of accurate information about $\omega$. We remark that other types of dynamical decoupling methods, e.g. XY-8 sequence (Appendix \ref{appendix-a}) and concatenated continuous driving \cite{Cai2012,Cai_2013,Stark2017} are also available to extract signal information and isolate the sensor from its noisy environment. The use of continuous drive is shown to be able to increase resolution of quantum NMR spectroscopy \cite{Aharon2019,Gefen2018,Cohen2019}. We also note that high resolution Hartmann-Hahn NV-NMR spectrometer was proposed in \cite{Vaknin2020} and was shown to be robust against magnetic field inhomogeneities. Moreover, the single drive that transfers Hartman - Hahn to a Qdyne type detection was proposed in \cite{Aharon2019}. In this work we extend these ideas to the high frequency domain together with the utilization of pulsed dynamical decoupling.
\begin{figure}[t]
  \includegraphics[width=86mm]{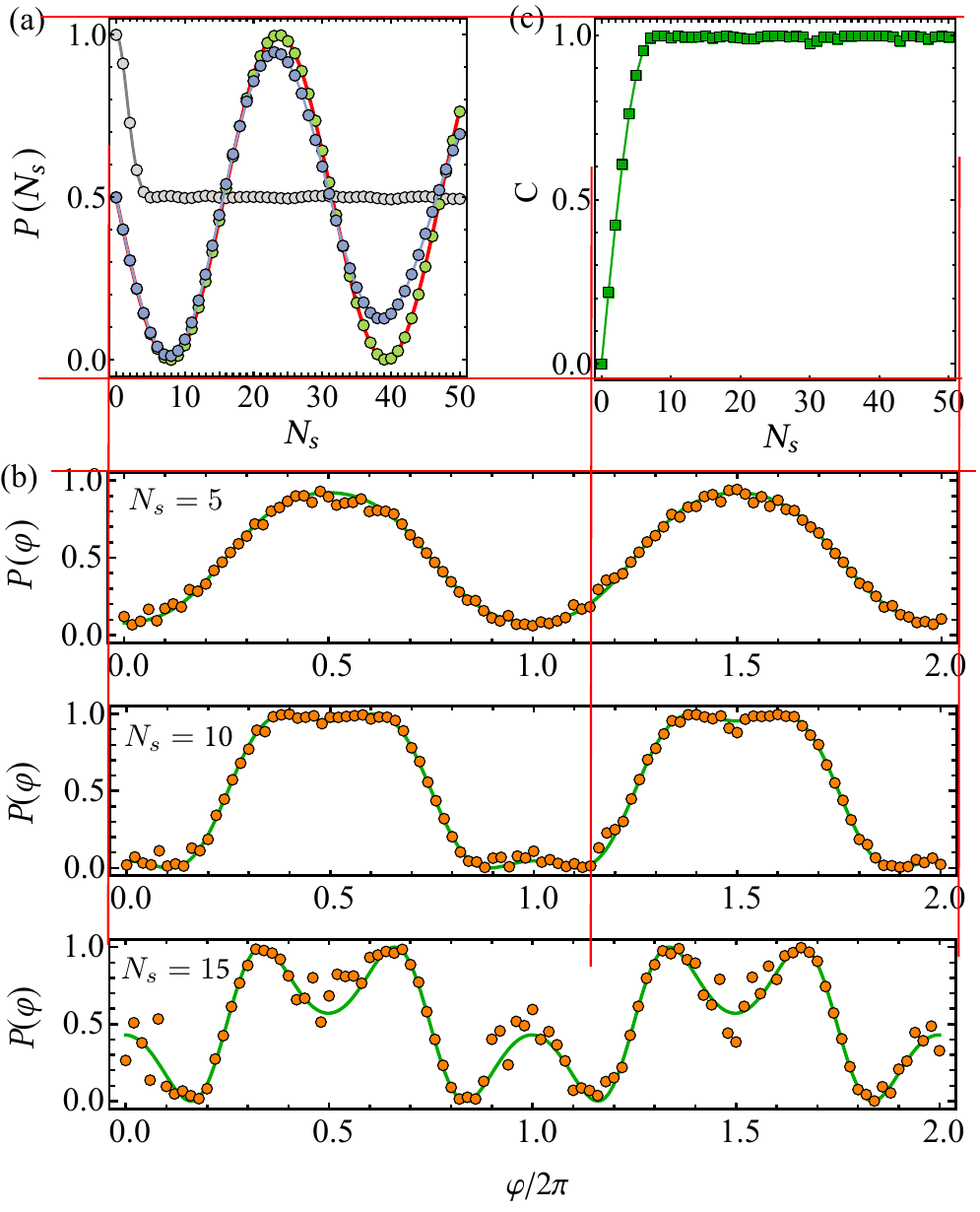}
  \caption{Population signal in a single measurement. (a) Population $P(N_s)=P(\varphi=0)|_{T_s=2N_s\tau}$ versus the number $N_s$ of applied CPMG sequences (green filled circle), obtained by numerically solving the quantum dynamics governed by the Hamiltonian in Eq.\eqref{eq:H_RF} as well as the decoupling sequences, agrees well with the red curve associated with the analytical result in Eq.\eqref{eq:signal}. As a comparison, we also simulate the same process in the presence of magnetic noise (blue filled circles) by averaging over 500 noise realizations. Moreover, a pure dephasing process starting from state $|+\rangle$, induced only by the magnetic noise, is depicted as well by calculating the population remaining in the initial state (gray filled circles).  (b) Population $P(\varphi)$ at a fixed number $N_s$ of applied CPMG sequences under single noise realization (orange filled circles), fitting well with the analytical result in Eq.\eqref{eq:signal} (green curve), shows an exact $2\pi$-periodicity versus the initial phase $\varphi$ of the target field. (c) Dependence of signal contrast $C$ on the number $N_s$ of applied CPMG sequences. The parameters are chosen as $k_s/2\pi=50$kHz, $\tau=0.5$us, and $\Delta/2\pi=1\text{MHz}+0.232\text{kHz}$. We model the magnetic noise $\delta B(t)$ as an O-U process with a correlation time $\tau_B=4$ms and a standard deviation $\Delta_B/2\pi=100$kHz.} \label{fig:populationCPMG}
\end{figure}

\section{Frequency-measurement protocol}

The protocol consists of many periodic successive measurements over the coherence time of the target field, see Fig.\ref{fig:model} (c). Each single measurement contains three steps, including an interaction time $T_s$ with the target field, an initialization and readout time $T_r$ of the sensor and an additional delay time $T_d$ to adjust the sampling rate. Hence the length of a single measurement is given by $T_L=T_s+T_r+T_d$, defining a sampling frequency $f_L=1/T_L$. In the $n$-th single measurement, we initialize the sensor in state $|1\rangle$, after an evolution time $T_s$ governed by the Hamiltonian in Eq.\eqref{eq:H_CPMG}, the final population of the sensor in superposition state $|+\rangle=(|0\rangle + |1\rangle)/\sqrt{2}$ is found to be \begin{equation}
  \label{eq:signal}
 P_n\equiv P(\varphi_n) =\sin^2[\Phi(\varphi_n)-\frac{\pi}{4}],
\end{equation}
where $\varphi_n  =\omega t_n+\varphi$ with $t_n=(n-1)T_L$ denoting the starting time of the $n$-th measurement and $\Phi(\varphi)=k_s T_s\text{sinc}(\delta T_s/2)\cos(\delta T_s/2+\varphi)/\pi$. We note that the phase $\varphi_n$ can be equivalently rewritten as $\varphi_n =\delta_L t_n+\varphi$, where $\delta_L=\omega-2\pi N_L f_L$ is the reduced signal frequency with $N_L$ an appropriate integer to guarantee the condition $|\delta_L/2\pi|<f_L/2$. The validity of Eq.\eqref{eq:signal} is testified by our exact numerical simulation governed by the Hamiltonian in Eq.\eqref{eq:H_RF} together with a train of CPMG sequences, see Fig.\ref{fig:populationCPMG} (a), where we also take into account the influence of slow magnetic noise as a comparison. The noise effect is described by an additional term $H_{\text{noise}}=\delta B(t)\sigma_z/2$ where the noise $\delta B(t)$ is phenomenologically modelled by a stochastic Ornstein-Uhlenbeck (O-U) process \cite{Gardiner2004}, which is reasonable for several quantum platforms, e.g. NV color centers \cite{Dobrovitski2009,deLange2010,Bermudez2011} or trapped ion systems \cite{Kotler2011} and can lead to a fast pure dephasing dynamics. It can be seen that the main effects of noise can be eliminated by dynamical decoupling sequences, resulting in a significantly extended coherence time of the sensor, which would play a significant role in the case of weak signal detection. On the other hand, $P(\varphi)$ at a fixed evolution time, even though with visible fluctuations in the presence of a single noise realization, exhibits an exact $2\pi$-periodicity versus the initial phase $\varphi$ of the target field, as shown in Fig.\ref{fig:populationCPMG} (b). We remark that this feature is crucial for us to correlate successive single measurements of the whole Qdyne measurement chain by precise timekeeping and extract the information on the target frequency beyond the limitation set by the sensor's coherence time \cite{Boss2017,Schmitt2017}. In order to find an appropriate number $N_s$ of CPMG sequences to be applied, we further define  the signal contrast as follows
\begin{equation}
    C=\frac{\text{Max}[P(\varphi)]-\text{Min}[P(\varphi)]}{\text{Max}[P(\varphi)]+\text{Min}[P(\varphi)]},
\end{equation}
 where Max (Min) represents the maximum (minimum) value of $P(\varphi)$ as a function of the signal phase $\varphi$. It can be seen from Fig.\ref{fig:populationCPMG} (c) that the contrast $C$ increases and would nearly saturates to 1 as $N_s$ grows. Thus, we can set the required sequence number in each single measurement run as the number $N_s=N_s^*$ at which $C$ begins to saturate.

State-selective fluorescence detection is an efficient method to readout the state of several quantum systems, such as trapped ions \cite{Keselman2011} and solid-state spins in diamond \cite{Hopper2018}, which defines a map $\mathcal{M}$ transforming the population signal $P_n$ into a random variable $z_n$ denoting the number of photons collected in the $n$-th experimental run
\begin{equation}
  \label{eq:photon}
  z_n=\mathcal{M}[P_n].
\end{equation}
As an ideal example, $z_n=\text{Bn}[P_n]$ represents a Bernoulli random process which takes the value $1$ with a probability of $P_n$ and the value $0$ with a probability of $1-P_n$. The Bernoulli process based mapping only involves quantum projection noise. More realistically, we assume that $z_n= \text{Pois}[\mu_0+(\mu_1-\mu_0)\text{Bn}[P_n]]$, or equivalently $z_n\sim P_n \text{Pois}[\mu_1]+(1-P_n)\text{Pois}[\mu_0]$, where $\text{Pois}[\mu_i]$, $i=0,1$, describes a Poissonian process with the mean value $\mu_i$, which is connected with the photon shot noise.
\begin{figure}[t]
  \includegraphics[width=84mm]{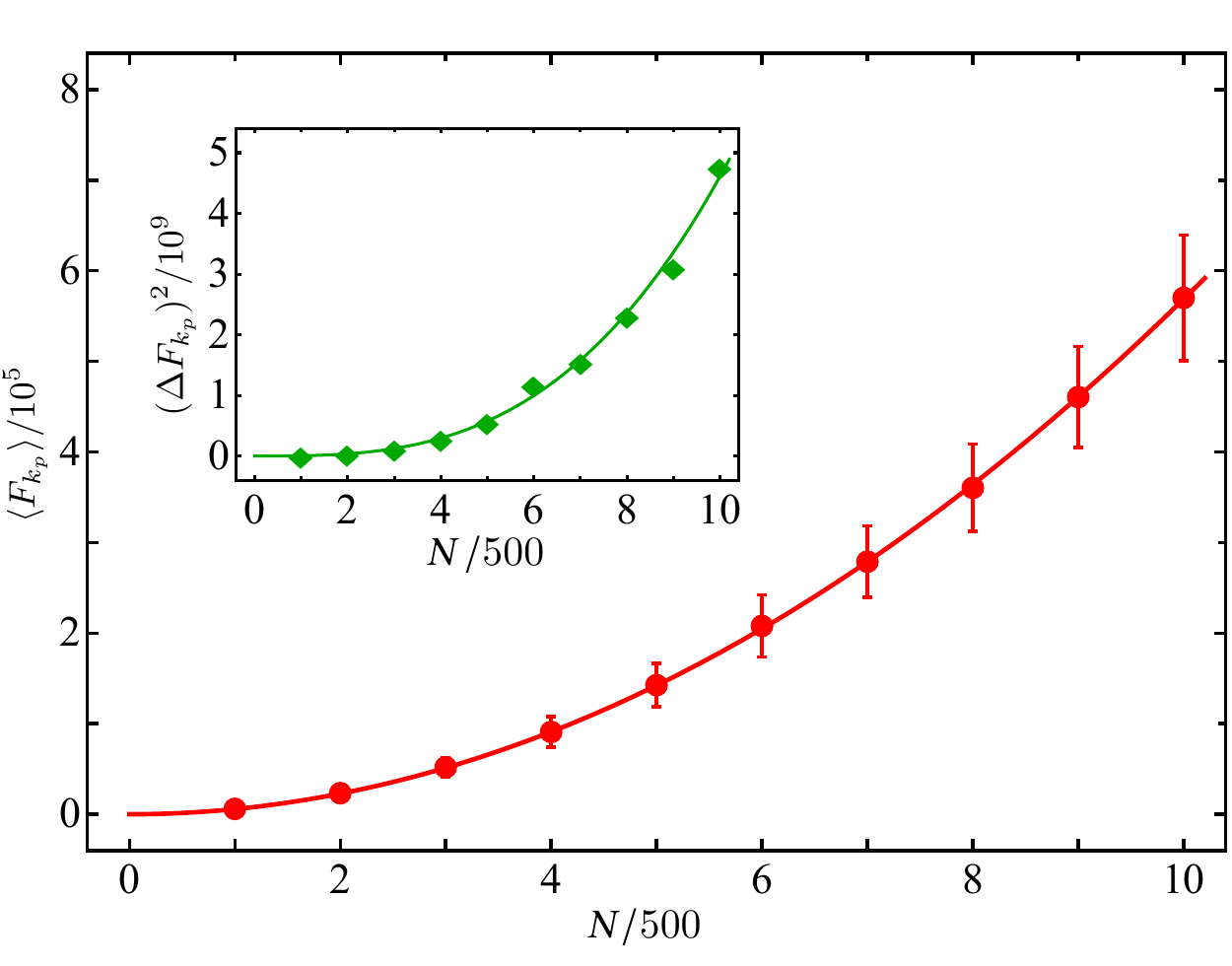}
  \caption{Dependence of average spectral density $\langle F_{{k}_p} \rangle$ at the signal peak and its fluctuation $(\Delta F_{{k}_p})^2$ on the total measurement runs $N$ in a Qdyne experiment chain. As an illustrative example, we simulate 500 different realizations of a stochastic process $z_n= \text{Pois}[0.1+\text{Bn}[P_n]]$ with $P_n=0.5+0.3\cos[0.3\pi (n-1)]$, $n=1,2,\cdots,N$, from which $\langle F_{{k}_p}\rangle$ (red dots) and $\Delta F_{{k}_p}$ (error bar) with $k_p=0.15N$ are calculated and fitted by $\langle F_{{k}_p}\rangle =0.0228 N^2$ (red line), $(\Delta F_{{k}_p})^2=0.0369 N^3$ (green line in the inset). We remark that both fitting results agree well with the analytical results given by Eq.\eqref{eq:SNR} as $\langle F_{{k}_p}\rangle \approx 0.0225 N^2$ and $(\Delta F_{{k}_p})^2\approx 0.0372 N^3$.}\label{fig:SNR}
\end{figure}

The information about the signal frequency is imparted onto a time trace of $N$ measurement outcomes $\{z\}_{n=1}^{N}$ at sampling times $\{t_n\}_{n=1}^{N}$. We extract the target frequency $\omega$ by making a discrete Fourier transform $\tilde{z}_k=\sum_{n=1}^N z_n e^{i2\pi n k/N}$, $k=0,1,\cdots,N-1$, which corresponds to the amplitude at the frequency component $f=k f_L/N$. The associated power spectrum $F_k=|\tilde{z}_k|^2$ is found to satisfy (Appendix \ref{appendix-b})
\begin{equation}
  \label{eq:FFT}
  \langle F_k\rangle=\left|\sum_{n=1}^N\langle z_n\rangle e^{i2\pi n\frac{k}{N}}\right|^2+\sum_{n=1}^N\mathrm{Var}[z_n],
\end{equation}
with $\langle z_n\rangle=\mu_0+(\mu_1-\mu_0)P_n$. It can be seen that the first term shows a peak at $\bar{\delta}_L=\delta_L N/(2\pi f_L)$ with a width $w=1/N$, to which we refer as the signal scaling as $N^2$, while the second term is a uniform part for all frequency components which scales as $N$. Therefore, when $N$ is large enough, the average spectral density $\langle F_k\rangle$ agrees with the spectral density of $\langle z_n\rangle$, which in principle allows us to extract $\delta_L$ from $\langle F_k\rangle$.
\begin{figure}[t]
  \includegraphics[width=84mm]{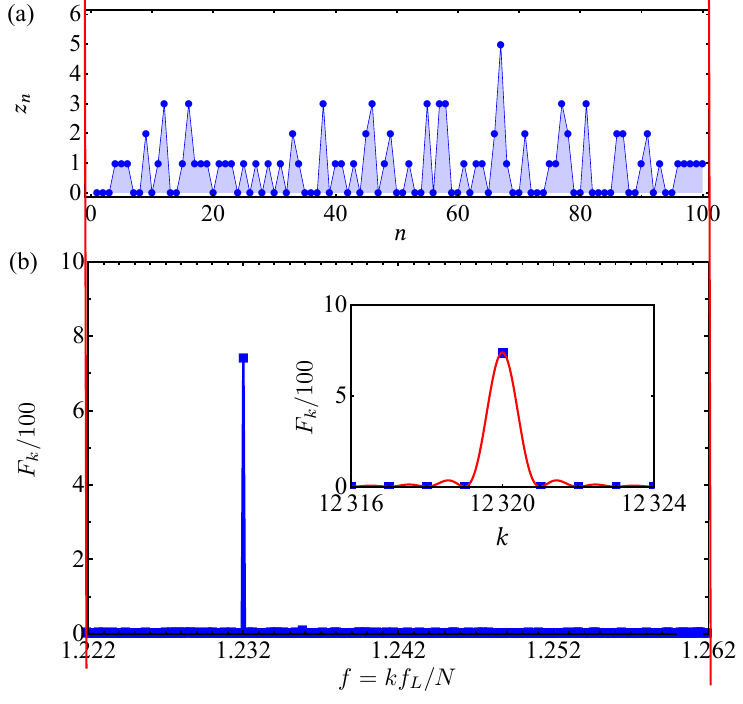}
  \caption{Numerical simulation of a Qdyne measurement chain with $N=10^5$ successive measurements. Each single measurement is simulated by numerically solving the quantum dynamics governed by the Hamiltonian in Eq.\eqref{eq:H_RF} together with a set of CPMG sequences in the presence of magnetic noise.  (a) First 100 outcomes of the measurement array $\{z_n\}_{n=1}^N$. Each blue dot denotes the number of emitted photons in a single experimental run generated by a map $z_n= \text{Pois}[0.7+0.3\text{Bn}[P_n]]$ where $P_n$ denotes the numerically solved population signal in state $|+\rangle$.   (b) Partial power spectrum $F_k$ of the measurement array derived by performing a discrete Fourier transformation of $\{z_n\}_{n=1}^N$. The inset shows an enlargement at the signal peak, which is fitted with the formula in Eq.\eqref{eq:fit} and yields $\bar{\delta}_L=12320$. Here, we choose the sampling frequency as $f_L=10$kHz, the parameters of the oscillating field as $\omega/2\pi=1801\text{MHz}+501.232\text{kHz}$, $k_s/2\pi=50$kHz, $\Delta/2\pi=1\text{MHz}+0.232\text{kHz}$, $\gamma=0$, the CPMG sequences as $\tau=0.5$us, $N_s^*=9$. The parameters of the O-U magnetic noise are $\tau_B=4$ms, $\Delta_B/2\pi=0.1$MHz. The result leads to an estimated frequency as $\delta_L/2\pi=\bar{\delta}_L f_L/N=1232.0$Hz. }\label{fig:qdyne}
\end{figure}
\section{Analysis of sensing performance}

The power spectrum $\{F_k\}_{k=0}^{N-1}$ derived from a Qdyne measurement chain is only a single realization of a stochastic process with the mean value $\langle F_k\rangle$. In order to verify the possibility of extracting $\delta_L$ from $F_k$, we demonstrate that the fluctuation of this stochastic process, defined by $(\Delta F_k)^2=\langle F_k\rangle^2-\langle F_k^2\rangle$, is negligible compared to the signal itself, i.e. $\langle F_k\rangle$ around its peak. Without loss of generality, by assuming a process $P_n=a+b\cos(\delta_L t+\varphi)$ with $a$, $b$ certain real parameters (Appendix \ref{appendix-b}), the signal and its fluctuation are found to be
\begin{equation}
  \begin{aligned}
    &\langle F_{{k}_p}\rangle \approx b^2(\mu_1-\mu_0)^2\beta N^2/4,\\
   & (\Delta F_{{k}_p})^2=\eta N^3,
  \end{aligned}\label{eq:SNR}
\end{equation}
where ${k}_p$ denotes the neighboring integers close to $\bar{\delta}_L=\delta_L N/(2\pi f_L)$, and the coefficients $\beta$, $\eta$ depend on the difference value ${k}_p-\bar{\delta}_L$. Based on the above result in Eq.\eqref{eq:SNR}, we find that the signal-to-noise ratio near the signal peak scales as $\mathcal{R}=\langle F_{{k}_p}\rangle/\Delta F_{{k}_p}\sim \mathcal{O}(\sqrt{N})\gg 1$, implying that $F_{{k}_p}$ obtained from a single stochastic realization agrees well with
$\langle F_{{k}_p}\rangle$ although accompanied by a relatively small fluctuation, and can thus be exploited to extract $\delta_L$. The scalings $\langle F_{{k}_p}\rangle\sim N^2$ and $(\Delta F_{{k}_p})^2\sim N^3$ are also verified by our exact numerical simulations, as shown in Fig.\ref{fig:SNR}. This analysis indeed shows that the proposed methods scales in a similar way to Qdyne \cite{Schmitt2017}.

Furthermore, we numerically simulate a complete Qdyne measurement process, each single measurement run of which is governed by the Hamiltonian in Eq.\eqref{eq:H_RF} together with longitudinal magnetic noise, see Fig.\ref{fig:qdyne}. In addition, we assume a relatively small state-dependent fluorescence, e.g. $\mu_0=0.7$, $\mu_1=1$, which is similar to the experimental readout efficiency of NV center in diamond. It can be seen that the Fourier spectrum $F_k$ shows an ultra-sharp peak at the reduced target frequency $\delta_L$, the width of which is $w\simeq \delta f\equiv f_L/N$. In order to accurately extract $\delta_L$, we use the following function with undetermined parameters $\{A,B,\bar{\gamma},\bar{\delta}_L\}$ (Appendix \ref{appendix-b})
\begin{equation}
  \label{eq:fit}
  F_k=A \frac{\cosh(\bar{\gamma}) -\cos\left[2\pi(k-\bar{\delta}_L)\right]}{\bar{\gamma}^2+4\pi^2(k-\bar{\delta}_L)^2}+B,
\end{equation}
to fit the neighborhood points of the signal peak, which enables us to determine $\delta_L$ with a precision $\sim w/\mathcal{R}\sim N^{-3/2}$. Note that $\bar{\gamma}$ is associated with the intrinsic linewidth $\gamma$ of the target field by the relation $\gamma=\bar{\gamma}f_L/N$.

Measurement precision of the signal frequency can also be analyzed from the perspective of quantum Fisher information (QFI). For the $n$-th measurement, the state population is given by $P_n=\sin^2(\Phi_n-\pi/4)$ with $\Phi_n=\Phi(\varphi_n)$ based on a measurement basis $|+\rangle\langle +|$, which is optimal with respect to the sensor's final state \cite{Schmitt2017}, thus the QFI of a single measurement with respect to the target frequency $\omega$ is given by
\begin{equation}
  \mathcal{I}_n(\omega)= \frac{|\partial P_n/\partial \Phi_n|^2}{P_n(1-P_n)}\left|\frac{\partial\Phi_n}{\partial \omega}\right|^2.
\end{equation}
Under the condition of $(\omega-\omega_0-\pi/\tau)T_s=\delta T_s\ll 1$, we obtain $\mathcal{I}_n(\omega)\approx 4 k_s^2T_s^2(t_{n}+T_s/2)^2\sin^2(\omega t_{n}+\varphi)/\pi^2$ and further the QFI of the whole Qdyne measurement chain as
\begin{equation}
\mathcal{I}(\omega)=\sum_{n=1}^{N}\mathcal{I}_n(\omega)\approx 2 k_s^2T_s^2T_L^2 N^3/(3\pi^2) \sim k_s^2T_s^2T_L^{-1}T^3
\end{equation}
for $N\gg 1$. On the other hand, the minimum frequency change that can be detected above the noise level satisfies the well-known quantum Cram\'er-Rao bound \cite{Braunstein1994}, i.e. $\delta\omega \geqslant 1/\sqrt{\mathcal{I}(\omega)}$. Therefore, the measurement precision scales as $\delta\omega\propto k_s^{-1}T_s^{-1}T_L^{\frac{1}{2}}T^{-\frac{3}{2}}$, which is consistent with the above analysis in the context of signal-to-noise ratio as in the Qdyne protocol \cite{Schmitt2017}.
\section{Conclusions \& Outlook}

We present an experimentally feasible scheme for precise spectroscopy of high-frequency oscillating fields using a single-qubit sensor, which is inaccessible for conventional quantum lock-in detection method. By approximately matching the two-level quantum sensor's energy splitting with the field frequency, the proposal effectively transforms the transverse field components to the RF range in a rotating frame. Thus, it provides a powerful method for quantum spectroscopy when the signal's oscillating period is much shorter than experimentally accessible time duration to implement spin-flipping $\pi$ pulses, which would significantly extend bandwidths of a single-qubit sensor. In combination with the dynamical decoupling technique, our scheme is robust against slow environmental noise acting on the sensor.  A further extension to more general control pulse sequences robust to pulse imperfections is possible and may increase the information that can be obtained from the oscillating fields \cite{Choi2020}. This result is expected to extend the application of quantum heterodyne detection in the high frequency regime.
\section{Acknowledgment}

 This work is supported by the National Natural Science Foundation of China (No. 11874024, 11690032), the Open Project Program of Wuhan National Laboratory for Optoelectronics (No. 2019WNLOKF002), the EU Flagship project AsteriQs (Grant No. 820394) and the ERC Synergy grant HyperQ (Grant No. 856432).

\textit{Note added.--} After the completion of this work, we became aware of the related work \cite{Meinel2020}

\section{Appendix}

\subsection{Derivation of the effective Hamiltonian in the toggling frame}
\label{appendix-a}
We consider a two-level quantum system acting as a single-qubit sensor, which is exposed to a fast oscillating field  and described by a total Hamiltonian as
\begin{equation}
 \label{eqs:H_t}
  H_t=\frac{\omega_0}{2}\sigma_z+\vec{b}\cdot \vec{\sigma}\cos(\omega t+\varphi_s)+\frac{\delta B(t)}{2}\sigma_z,
\end{equation}
where $\delta B(t)$ describes slow environmental noise with a correlation time $\tau_B$ and a standard deviation denoted by $\Delta_B$. We remark that transverse noise in the $x$-$y$ plane is largely suppressed by the energy splitting $\omega_0$ of the qubit sensor. By moving to the interaction picture with respect to $H_0=(\omega_0/2)\sigma_z$, we obtain
\begin{eqnarray}
\label{eqs:H_RF}
  \nonumber  \mathcal{H}_t&=&e^{iH_0t}(H_t-H_0)e^{-iH_0t}\\
  \nonumber               &=&\frac{\delta B(t)}{2}\sigma_z+b_z\cos(\omega t+\varphi_s)\sigma_z\\
  \nonumber               &+&k_s\cos(\omega t+\varphi_s)\left(\sigma_+e^{i\omega_0t-i\theta}+\sigma_-e^{-i\omega_0t+i\theta}\right)\\
  \nonumber              &\approx&\frac{\delta B(t)}{2}\sigma_z+\frac{k_s}{2}\cos(\Delta t+\varphi)\sigma_x\\
                          &+&\frac{k_s}{2}\sin(\Delta t+\varphi)\sigma_y.
\end{eqnarray}
under the conditions of $k_s=\sqrt{b_x^2+b_y^2}\ll |\Delta| \ll \omega+\omega_0$ and $b_z\ll\omega$, wherein $\Delta=\omega-\omega_0$ and $\tan(\theta)=b_y/b_x$. More accurately, $\Delta$ can be defined as $\Delta=\omega-\omega_d$, where $\omega_d$ represents the frequency of external control fields, i.e. spin-flipping $\pi$-pulses. In this case, we obtain an additional term $\mathcal{H}^\prime=(\delta_d/2)\sigma_z$, $\delta_d=\omega_0-\omega_d$ to Eq.\eqref{eqs:H_RF} in the interaction picture with respect to $H_0'=(\omega_d/2)\sigma_z$. We remark that a small detuning of $\delta_d$ can be eliminated by dynamical decoupling sequences, which makes the proposal in the main text more feasible and robust in experiments.
\textit{CPMG sequence.--}
In order to extract signal information in Eq.\eqref{eqs:H_RF} and prolong the sensor's coherence time in a single measurement, we apply a train of CPMG sequences i.e. $(\tau$-$\pi$-$\tau$-$\pi)^{N_s}$ along the $\hat{x}$-direction, the control Hamiltonian of which in the lab frame reads $H_c(t)=\Omega(t)\cos(\omega_0t)\sigma_x$ and transforms as $\mathcal{H}_c\approx[\Omega(t)/2]\sigma_x$ in the interaction picture with respect to $H_0=(\omega_0/2)\sigma_z$ ($|\Omega(t)|\ll\omega_0$). Here, $\Omega(t)$, characterizing a train of sharp $\pi$ pulses, can be approximated as a set of periodic Dirac delta functions with a equidistant spacing of $\tau$. By moving to the interaction with respect to these pulses, the Hamiltonian in Eq.\eqref{eqs:H_RF} in the toggling frame \cite{Choi2020} is found to be
\begin{eqnarray}
  \label{eqs:H_eff_cpmg_origin}
\nonumber
  \mathcal{H}(\varphi)&=&\frac{k_s}{2}\cos(\Delta t+\varphi)\sigma_x+f(t)\frac{\delta B(t)}{2}\sigma_z\\
  &+&f(t)\frac{k_s}{2}\sin(\Delta t+\varphi)\sigma_y,
\end{eqnarray}
where $f(t)$ is a square wave modulation function arising from periodic $\pi$-pulses. More precisely, $f(t)=1$ if $0\leqslant \text{mod}(\omega_s t,2\pi)<\pi$ otherwise $f(t)=-1$ with $\omega_s=\pi/\tau$, which can be expanded as $f(t)=\sum_{n=\text{odd}}{(4/n\pi)}\sin(n\omega_s t)$. Under the conditions of $\omega_s\approx \Delta \gg \{k_s,\Delta_B\}$ and $\tau\ll \tau_B$, we obtain (i.e. Eq.\textcolor{red}{2} in the main text)
\begin{equation}
  \label{eqs:H_eff_cpmg}
  \mathcal{H}(\varphi)\approx\frac{k_s}{\pi}\cos(\delta t+\varphi)\sigma_y,
\end{equation}
where $\delta=\Delta-\omega_s\approx 0$. If we initialize the sensor in state $|1\rangle$ and let it evolve under the Hamiltonian in Eq.\eqref{eqs:H_eff_cpmg} for time $T_s$, the population in the state $|+\rangle=(|0\rangle + |1\rangle)/\sqrt{2}$ is found to be
\begin{eqnarray}
  \label{eqs:population}
  \nonumber P(\varphi)&=&\left|\langle+|e^{-i\int_0^{T_s}\frac{k_s}{\pi}\cos(\delta t+\varphi)dt \sigma_y}|1\rangle\right|^2\\
  \nonumber           &=&\left|\langle+|e^{-i\Phi(\varphi)\sigma_y}|1\rangle\right|^2\\
  &=&\sin^2[\Phi(\varphi)-\frac{\pi}{4}],
\end{eqnarray}
where $\Phi(\varphi)=k_s T_s\text{sinc}(\delta T_s/2)\cos(\delta T_s/2+\varphi)/\pi$ \cite{Schmitt2017}.

In order to evaluate the noise influence to the next order, we first investigate a periodic Hamiltonian of the following form
\begin{equation}
  \label{eq:periodic-H}
    H(t)=\sum_{n>0}\left[\cos(n\omega t)\hat{A}_n+\sin(n\omega t)\hat{B}_n\right]+\hat{C}.
  \end{equation}
By invoking the formula of the first two orders of Magnus series \cite{Blanes2009}, the effective Hamiltonian can be calculated by
  \begin{equation}
    H_{\text{eff}}=\frac{1}{T}\int_0^TH(t)dt-\frac{i}{2T}\int_0^T\mathrm{d}t_1\int_0^{t_1}\mathrm{d}t_2[H(t_1),H(t_2)],
  \end{equation}
where $T=2\pi/\omega$. By plugging Eq.\eqref{eq:periodic-H} into the above formula and performing direct integration, we obtain
\begin{equation}
 H_{\text{eff}}=\hat{C}+\sum_{n>0}\frac{i}{2n\omega}[\hat{A}_n,\hat{B}_n]-\sum_{n>0}\frac{i}{n\omega}[\hat{C},\hat{B}_n].
\end{equation}
According to this result, the effective Hamiltonian corresponding to the Hamiltonian in Eq.\eqref{eqs:H_eff_cpmg_origin} reads
\begin{eqnarray}
\nonumber  \mathcal{H}(\varphi)&\approx&\frac{k_s}{\pi}\cos(\delta t+\varphi)\sigma_y\\
                               &+&\frac{i}{2\omega_s}\left[\frac{k_s}{2}\cos(\delta t+\varphi)\sigma_x,\frac{2\delta B(t)}{\pi}\sigma_z\right]\\
  \nonumber &=&\frac{k_s}{\pi}\cos(\delta t+\varphi)\left[1+\frac{\delta B(t)}{\omega_s}\right]\sigma_y,
\end{eqnarray}
where other components complementary to $\sigma_y$, i.e. $\sigma_{x,z}$, are omitted due to the fact that $k_s\gg \{k_s^2/\omega_s, k_s\delta B(t)/\omega_s\}$. From this effective Hamiltonian, it can be seen that the noise effect is more evident at $\varphi=n\pi$, $n\in\mathbb{Z}$ as compared with the situation of $\varphi=n\pi+\pi/2$.

\begin{figure}[t]
  \includegraphics[width=86mm]{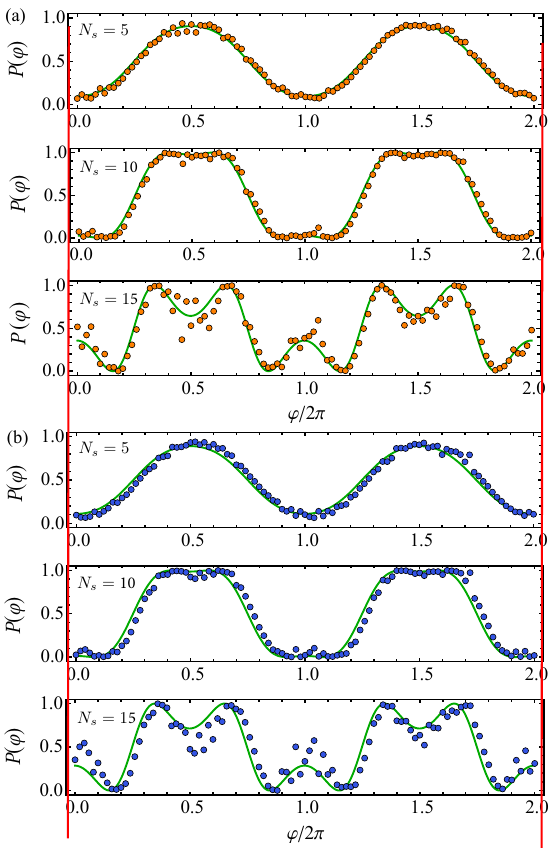}
  \caption{Influence of finite-width $\pi$ pulses. Population $P(\varphi)$ at fixed number $N_s$ of applied CPMG sequences (filled circles) is obtained by numerically simulating the quantum dynamics governed by the Hamiltonian in Eq.\eqref{eq:H_RF} accompanied by the decoupling sequences and magnetic noise. The widths of single applied $\pi$ pulse are chosen as $t_p=25$ns (a) and $t_p=50$ns (b) respectively. As we see, finite-width $\pi$ pulses lead to a slight overall shift of $P(\varphi)$ versus $\varphi$ as compared to the case of instantaneous $\pi$ pulses (green curves). Here, we choose $k_s/2\pi=50$kHz, $\Delta/2\pi=1\text{MHz}+0.232\text{kHz}$ and the free evolution time between adjacent $\pi$ pulses as $\tau=\tau_s-t_p$ with $\tau_s=0.5$us. The magnetic noise $\delta B(t)$ is modelled as an O-U process with a correlation time $\tau_B=4$ms and a standard deviation $\Delta_B/2\pi=100$kHz.} \label{fig:finite-pulse}
\end{figure}

On the other hand, instantaneous $\pi$ pulses in the above derivations are assumed for simplicity due to the fact that $\pi/\Delta \gg t_p $ with $t_p$ the single $\pi$-pulse width. More accurately, finite-width $\pi$ pulses in real experiments should result in small corrections to the population signal in Eq.\eqref{eqs:population}, e.g. a slight overall shift of $P(\varphi)$ versus $\varphi$, as shown in Fig.\ref{fig:finite-pulse}. However, we remark that the exact $2\pi$-periodicity of $P(\varphi)$ versus $\varphi$ is not altered by such experimentally feasible finite-width $\pi$ pulses, which thus still allows us to correlate successive measurements and extract the exact frequency information about the target field.

\vspace{0.1cm}

\textit{XY-8 sequence.--}
As an alternative efficient method to decouple environmental noise and measure the effective signal field in Eq.\eqref{eqs:H_RF}, we can also use a train of XY-8 sequences, i.e. $(\tau$-$\pi_x$-$\tau$-$\pi_y$-$\tau$-$\pi_x$-$\tau$-$\pi_y$-$\tau$-$\pi_y$-$\tau$-$\pi_x$-$\tau$-$\pi_y$-$\tau$-$\pi_x)^{N_s}$. By moving to the interaction picture with respect to this control field, we similarly obtain an effective Hamiltonian in the toggling frame as
\begin{eqnarray}
  \label{eqs:H_XY-8}
  \nonumber
  \widetilde{\mathcal{H}}(\varphi)=h(t)\frac{\delta B(t)}{2}\sigma_z&+f(t)\frac{k_s}{2}\cos(\Delta t+\varphi)\sigma_x\\
  &+g(t)\frac{k_s}{2}\sin(\Delta t+\varphi)\sigma_y,
\end{eqnarray}
where $f(t)$, $g(t)$ and $h(t)$ are modulation functions associated with the periodic XY-8 sequences. The Fourier expansion coefficients of these functions, corresponding to the harmonics $\cos(n\tilde{\omega}_s t)$, $\tilde{\omega}_s=\pi/(4\tau)$, $n\in\mathbb{Z}$, can be obtained as follows
  \begin{align}
    & f_n=\frac{4}{n\pi}\left[\sin(\frac{3n\pi}{8})-\sin(\frac{7n\pi}{8})\right], \\
    & g_n=\frac{4}{n\pi}\left[\sin(\frac{n\pi}{8})-\sin(\frac{5n\pi}{8})\right],  \\
    & h_n=\frac{4}{n\pi}\left[\sin(\frac{n\pi}{8})-\sin(\frac{3n\pi}{8})+\sin(\frac{5n\pi}{8})-\sin(\frac{7n\pi}{8})\right].
  \end{align}
In this case, we require $\tilde{\delta}=\Delta-\tilde{\omega}_s\approx 0$. In a similar way we derive Eq.\eqref{eqs:H_eff_cpmg}, the Hamiltonian in Eq. \eqref{eqs:H_XY-8} can be approximated as
\begin{equation}
  \widetilde{\mathcal{H}}(\varphi)=\frac{k_s}{4}f_1\left[\cos(\tilde{\delta} t+\varphi)\sigma_x-\sin(\tilde{\delta} t+\varphi)\sigma_y\right],
\end{equation}
under the conditions of $\tilde{\omega}_s\gg \{k_s, \tilde{\delta},\Delta_B/4\}$ and $\tau\ll\tau_B$, where the denominator factor of $\Delta_B/4$ results from the fact that $h_n$ takes nonzero values only at $n=4m$, $m\in\mathbb{Z}^+$. If the sensor is initialized in the state $|+_y\rangle$, the final population in the state $|0\rangle$ after an evolution time $t$ is given by
\begin{eqnarray}
    \widetilde{P}(\varphi) &=&\left|\left\langle 0\right|e^{i\frac{\tilde{\delta}}{2}\sigma_z t}e^{-i(\frac{\tilde{\delta}}{2}\sigma_z+\frac{g}{2}\cos\varphi \sigma_x -\frac{g}{2}\sin\varphi \sigma_y) t}\left|+_y\right\rangle\right|^2\\ \nonumber
    &=&\frac{1}{2}\left[1+\frac{g}{\Omega}\cos(\varphi)\sin(\Phi)-\frac{2g\tilde{\delta}}{\Omega^2}\sin(\varphi)\sin^2(\frac{\Phi}{2})\right],
\end{eqnarray}
where $g=k_s f_1/2$, $\Omega=\sqrt{g^2+\tilde{\delta}^2}$ and $\Phi=\Omega t$.

\subsection{Fourier analysis of the Qdyne measurement}
\label{appendix-b}
The state population in Eq.\eqref{eqs:population} for the $j$-th measurement (based on CPMG-sequence) takes the form
\begin{equation}
  P_j=c+d\sin(\eta\cos(\omega t_{j-1}+\varphi+\delta T_s/2)),
\end{equation}
where $\eta=k_s T_s\text{sinc}(\delta T_s/2)/\pi$ and $c$, $d$ are some real parameters. Provided that $\eta<2$, $P_j$ can be approximated as
\begin{equation}
  P_j\approx c+2d\mathcal{J}_1(\eta)\cos(\omega t_{j-1}+\varphi+\delta T_s/2),
\end{equation}
based on $\sin(\eta\cos\theta)=2\sum_{n=0}^{\infty}(-1)^n\mathcal{J}_{2n+1}(\eta)\cos(2n+1)\theta$. While in the case of XY-8 sequence, we have
\begin{equation}
P_j\approx [\tilde{c}+\tilde{d} \cos(\varphi_j+\tilde{\theta})],
\end{equation}
where $\tilde{c}$, $\tilde{d}$ and $\tilde{\theta}$ are parameters depending on $g$, $\tilde{\delta}$ and $\Phi$. Therefore, without loss of generality, we consider an example of the form $P_n=a+b\cos(2\pi f t_n+\theta)e^{-\gamma t_n}$ in the following discussion, where $\gamma$ denotes the intrinsic linewidth of the target field. By assuming a photon emission process as $z_n\sim P_n\text{Pois}[\mu_1]+(1-P_n)\text{Pois}[\mu_0]$, we find that
\begin{eqnarray}
 \langle z_n\rangle&=&a_z+b_z\cos(2\pi f t_n+\theta)e^{-\gamma t_n},\\
\text{Var}[z_n]&=&\mu_0^2+(\mu_1^2-\mu_0^2)P_n+\langle z_n\rangle(1-\langle z_n\rangle),
\end{eqnarray}
where $a_z=\mu_0+a(\mu_1-\mu_0)$ and $b_z=b(\mu_1-\mu_0)$. Moreover, the power spectrum of $\{z_n\}_{n=1}^N$ is given by
\begin{equation}
  F_k=\sum_{n,m=1}^N z_n z_m e^{i2\pi(n-m)\frac{k}{N}}, \quad k\in (-N/2,N/2].
\end{equation}
Under the assumption that $\langle z_n z_m\rangle = \langle   z_n\rangle\langle z_m\rangle$ for $m\neq n$, the average of $F_k$ is found to be
\begin{eqnarray}
\nonumber
  \langle F_k \rangle &=&\left|\sum_{n=1}^N \left[a_z+\frac{b_z}{2}\left(e^{-i2\pi n\frac{\bar{f}}{N}-i\theta-n\frac{\bar{\gamma}}{N}}+\text{c.c.}\right)\right] e^{i2\pi n\frac{k}{N}}\right|^2\\
  &+&\sum_{n=1}^N\text{Var}(z_n),
\end{eqnarray}
with $\bar{f}=f N/f_L$ and $\bar{\gamma}=\gamma N/f_L$. In order to calculate $\langle F_k\rangle$, we define the following functions as
\begin{align}
  f_{\alpha}(x)&=\frac{1}{N}\sum_{n=1}^N e^{i2\pi n\frac{x}{N}-n\frac{\bar{\gamma}}{N}},\\
  f_{\beta}(x)&=\left|f_{\alpha}(x)\right|^2\\\nonumber
  &=\frac{1}{N^2}\frac{\cosh(\bar{\gamma})-\cos(2\pi x)}{\cosh(\frac{\bar{\gamma}}{N})-\cos(2\pi \frac{x}{N})}\\\nonumber
  & \approx \frac{2\left[\cosh(\bar{\gamma})-\cos(2\pi x)\right]}{\bar{\gamma}^2+4\pi^2 x^2}, \quad x/N\ll1, \bar{\gamma}/N\ll 1.
  \end{align}
Without loss of generality, we assume that $\bar{f}\gg 1$, $\frac{N}{2}-\bar{f}\gg 1$ and $0< k \leqslant N/2$. Thus, we can get
\begin{align}
\sigma&=\sum_{n}\text{Var}(z_n)\\\nonumber
&\approx \left[\mu_0^2+a(\mu_1^2-\mu_0^2)+a_z(1-a_z)\right]N-b_z^2\frac{\sinh(\bar{\gamma})}{2\sinh(\frac{\bar{\gamma}}{N})}\\\nonumber
&\sim \mathcal{O}(N),
\end{align}
and (i.e. Eq.\textcolor{red}{8} in the main text)
\begin{equation}
\langle F_{\bar{k}} \rangle\approx \frac{b_z^2}{4}\beta N^2 \sim \mathcal{O}(N^2),
\end{equation}
where $\beta=f_{\beta}(k_p-\bar{f})$ and $k_p$ denotes the neighboring integers of $\bar{f}$. Similarly the fluctuation of $F_k$ is given by
\begin{equation}
\left(\Delta F_k\right)^2\approx 2\sigma\langle F_k\rangle +F^{(3)}(k),
\end{equation}
where
\begin{eqnarray}
   \nonumber F^{(3)}&&(k)=\sum_{n,m,q}^N \left(\langle z_n^2\rangle-\langle z_n\rangle^2\right)\langle z_m\rangle \langle z_q\rangle \\
    &&\times \left(e^{i2\pi(2n-m-q)k/N}
    +e^{i 2\pi (m+q-2n)k/N}\right).
\end{eqnarray}
For $k={k}_p$, we can obtain
\begin{equation}
    F^{(3)}(k_p)\approx \frac{b_z^4}{8}\text{Re}\left[f_{\alpha}^2(\bar{f}-k_p)f_{\alpha}(2 k_p-2\bar{f})\right]N^3,
\end{equation}
which further leads to that (i.e. Eq.\textcolor{red}{8} in the main text)
\begin{equation}
  (\Delta F_{k_p})^2\approx \eta N^3,
\end{equation}
with
\begin{align}
\nonumber
\eta=\frac{b_z^2}{2}\beta \left[\mu_0^2+a(\mu_1^2-\mu_0^2)
+a_z(1-a_z)-\frac{b_z^2}{2}\right]\\+\frac{b_z^4}{8}\text{Re}\left[f_{\alpha}^2(\bar{f}-k_p)f_{\alpha}(2 k_p-2\bar{f})\right].
\end{align}

\bibliography{reference}
\end{document}